# X-ray microanalysis in STEM of short-term physico-chemical reactions at bioactive glass particles / biological fluids interface. Determination of O/Si atomic ratios.

V. BANCHET[*1], E. JALLOT[2], J. MICHEL[1], L. WORTHAM[1], D. LAURENT-MAQUIN[1] and G. BALOSSIER[1]

[1] INSERM ERM 0203, Laboratoire de Microscopie Electronique, 21 rue Clément Ader, 51685 REIMS Cedex 2, FRANCE

[2] Laboratoire de Physique Corpusculaire de Clermont-Ferrand CNRS/IN2P3 UMR 6533, 24 avenue des Landais, 63177 AUBIERE Cedex FRANCE

**Running head :** O/Si ratios during physico-chemical reactions at bioactive glass periphery.

**\* Corresponding author :**

BANCHET Vincent, INSERM ERM 0203, Laboratoire de Microscopie Electronique,

21 rue Clément Ader, 51685 REIMS Cedex 2, France

**Tel :** 33 (0)326050750

**Fax :** 33 (0)326051900

**E-mail :** vincent.banchet@univ-reims.fr

Abstract

Short-term physico-chemical reactions at the interface between bioactive glass particles and biological fluids are studied and we focus our attention on the measurements of O/Si atomic ratio. The studied bioactive glass is in the $SiO_2$-$Na_2O$-$CaO$-$P_2O_5$-$K_2O$-$Al_2O_3$-$MgO$ system. The elemental analysis is performed at the submicrometer scale by STEM associated with EDXS and EELS. We previously developed an EDXS quantification method based on the ratio method and taking into account local absorption corrections. In this way, we use EELS data to determine, by an iterative process, the local mass thickness which is an essential parameter to correct absorption in EDXS spectra. After different delays of immersion of bioactive glass particles in a simulated biological solution, results show the formation of different surface layers at the bioactive glass periphery. Before one day of immersion, we observe the presence of an already shown (Si,O,Al) rich layer at the periphery. In this paper, we demonstrate that a thin "electron dense" (Si,O) layer is formed on top of the (Si,O,Al) layer. In this (Si,O) layer, depleted in aluminium, we point out an increase of oxygen weight concentration which can be interpreted by the presence of $Si(OH)_4$ groups, that permit the formation of a (Ca,P) layer. Aluminium plays a role in the glass solubility and may inhibit apatite nucleation. After the beginning of the (Ca,P) layer formation, the size of the "electron dense" (Si,O) layer decreases and tends to disappear. After two days of immersion, the (Ca,P) layer grows in thickness and leads to apatite precipitation.

Introduction

Bioactive glasses, glass ceramics and calcium phosphate ceramics have received much attention for hard tissue replacement applications because they can form a direct bond with bone.[1] This property is known as bioactivity. A bioactive fixation is defined as « an interfacial bonding of an implant to tissue by means of formation of a biologically active hydroxyapatite layer on the implant surface ».[2] Hench established the relationship between glass composition and bioactive properties.[3] These bioactive materials can be used as coating on metallic prosthesis or as bone fillers,[4] but the composition has to be optimized to give a suitable compromise between bioactivity and solubility.[5] When exposed to biological fluids, bioactive glasses undergo a serie of physico-chemical reactions at the periphery, resulting in the formation of a silica gel and a calcium phosphate layer on their surfaces.[6, 7, 8]

However, the formation of these layers is not explained in detail because of the complexity of short-term events that happen at the interface. Knowledge of the elemental distribution at the bioactive glass periphery is important to understand the physico-chemical mechanisms involved in forming the Si gel and the (Ca,P) layer. This knowledge requires analysis at the submicrometer scale. This study was performed by scanning transmission electron microscopy (STEM) associated with energy-dispersive X-ray spectroscopy (EDXS) and with electron energy loss spectroscopy (EELS) thanks to a dedicated light elements quantification process.[9]

This study concerns a bioactive glass in the $SiO_2$-$Na_2O$-$CaO$-$P_2O_5$-$K_2O$-$Al_2O_3$-$MgO$ system. The silicon oxidation degree can play an important role in the regulation and the formation of physico-chemical processes, notably at the interface between a Si rich layer and the (Ca,P) layer. In this paper, we paid attention on this interface by measuring locally the O/Si atomic ratio, by means of X-rays microanalysis at the periphery of bioactive glass samples immersed for different delays (0, 12, 24 and 96 h) in a simulated biological fluid. The results will permit the evaluation of the silicon oxidation in the Si rich layer. On the other hand, aluminium may inhibit apatite precipitation or reduces glass matrix dissolution. In this paper, we also take into account the role of Al during short-term physico-chemical reactions.

Materials and Methods

*Composition of the bioactive glass*

The composition of the bioactive glass (referenced as A9) is : 50% $SiO_2$, 20% $Na_2O$, 16% CaO, 6% $P_2O_5$, 5% $K_2O$, 2% $Al_2O_3$ and 1% MgO (% weight). Addition of $Al_2O_3$ may be used to control the solubility of the glass, but this addition may inhibit the bone bonding.[5, 10] Greenspan D.C.[11] showed that 3% $Al_2O_3$ added to a bioactive glass inhibits bone bonding. In our case 2% of $Al_2O_3$ was added and the bioactive glass is in the $SiO_2$-$Na_2O$-CaO-$P_2O_5$-$K_2O$-$Al_2O_3$-MgO system.

*Samples preparation*

The bioactive glass was obtained by melting the components at 1350 °C. Then, the glass was cast, crushed, and transformed into powder of grain size under 40 µm in diameter. The glass powder (2 mg) was immersed at 37 °C for different delays (0, 12, 24 and 96 h) in 1 mL of a standard Dulbecco's Modified Eagle Medium (DMEM).[12] DMEM is a simulated body fluid and contains the following ingredients (mg/L) : 6400 NaCl, 400 KCl, 200 $CaCl_2$, 200 $MgSO_4$ 7 $H_2O$, 124 $NaH_2PO_4$, and 3700 $NaHCO_3$. Then the bioactive glass powder is embedded in resin (AGAR, Essex, England). Thin sections of 90 nanometers nominal thickness are prepared by means of a FC 4E Reichert Young ultramicrotome. The sections are placed on a copper grid (200 Mesh). Sections were coated with a conductive layer of carbon in a sputter coater to avoid charging effects.

*Analysis materials*

Our experiences are carried out using a scanning transmission electron microscope (STEM) (Philips CM30). The microscope is fitted with an energy dispersive X-ray spectrometer (EDAX 30 mm² Si(Li) R-SUTW detector) and with a parallel electron energy loss spectrometer (GATAN model 666) placed under the STEM column. Analyses are carried out using a beryllium specimen holder with 30°-tilt. The electron probe diameter is around 13 nanometers. EDXS spectra are acquired at an accelerating voltage of 100 kV, with an energy resolution of 10 eV per channel and an electron dose of approximately 3 x 10$^6$ e$^-$/nm². EELS experiments are performed at an accelerating voltage of 250 kV with an energy dispersion of 0.1 eV per channel.

*Methods*

Energy-dispersive X-ray spectrometry, EDXS, is a common way to perform elemental analysis in a transmission electron microscope.[13, 14] Recent developments of thin-window or windowless detectors allow the low Z elements (like boron, carbon, nitrogen, oxygen, fluorine,…) detection.[15] Nevertheless, quantitative analysis of these elements is problematic due to the absorption of the low energy X-rays in the sample itself.[16] Using the quantification classical ratio method (also known as Cliff and Lorimer method)[17] and without taking into account of absorption phenomena leads to underestimate the concentrations of light elements. Moreover the concentrations are related to each others. In this way, absorption of low energy X-ray influences not only the concentrations of light elements but also the concentrations of all other elements. Mostly absorption correction requires the knowledge of the local mass thickness.[14, 18] In the case of complex samples, heterogeneous in composition, density and thickness, like bioactive glasses at different steps of dissolution in contact with biological fluids, it is not possible to use the nominal value of section thickness or a global value of sample density to correct absorption. It is then necessary to determine experimentally the local mass thickness. We use EELS data to obtain experimentally the local mass thickness. More precisely, by combining EELS measured relative specimen thickness, and X-rays characteristic peaks intensities, we determine with an iterative process the local mass thickness. More details are given in a previous publication dedicated to low Z elemental quantification in STEM.[9] Thus, we can obtain weight concentrations of all elements of the bioactive glass, including oxygen. In order to evaluate the O/Si and Ca/P atomic ratios across the different layers, we performed line spectra acquisition profiles.

Results

As a first check, we ensure that no irradiation damage occurs on the specimens during EDXS acquisition in our experimental conditions (electron dose of 3 x $10^6$ e$^-$/nm²). In this way, we measured X-rays characteristic peak intensities for different electron doses up to $10^7$ e$^-$/nm². We do not detect any significant elemental concentration variations in these typical analysis conditions.

Weight concentrations variations in O, Si, Al, Ca and P at the bioactive glass / biological fluids interface were studied by EDXS at each time of exposure to DMEM. For each analysed area, we acquired an EELS spectrum in the low-loss region to obtain the local relative specimen thickness in order to perform quantitative analysis as explained in the previous section.

For each delay of immersion, differents weight concentrations profiles have been acquired (7 profiles by sample and 3 samples were used). Results are characterised by a great reproducibility with a standard deviation lower than 5 %. Thus, we present only one typical profile for each delay in the following sections. These profiles represent the average behaviour of all glass particles.

*Bioactive glass periphery in the original state*

Figure 1 presents an inverted darkfield STEM image of the bioactive glass. We can observe in the area of the bioactive glass that the mass thickness is rather inhomogeneous. Figure 2 shows the distribution of O, Si, Al, Mg, Na, K, Ca and P weight concentrations across the periphery of bioactive glass particles. We can conclude that the distribution of the absorption corrected weight concentrations is homogeneous in the bioactive glass particles.

These experimental weight concentrations obtained are close to the theoretical values. Finally, we do not observe any formation of specific layers at the periphery.

*Bioactive glass periphery after 12 h of immersion in DMEM*

The inverted darkfield STEM image shows a part of the bioactive glass after 12 h of immersion in DMEM (figure 3). Elemental weight concentration profiles of Si, O, Al, Ca and P reveal different zones from the center to the periphery of the bioactive glass (figure 4). We observe the formation of a (Si,O,Al) layer at the bioactive glass/biological fluids interface. We also observe an additionnal (Si,O) thin layer (a few hundred of nanometers) formed on top of the (Si,O,Al) layer. This (Si,O) layer appears as an "electron dense" layer in the inverted darkfield STEM image (figure 3). Moreover, in this layer, oxygen weight concentration is higher than in the (Si,O,Al) layer. These two characteristics lead us to distinguish the "electron dense" (Si,O) from the (Si,O,Al) layer. We also see that, in the "electron dense" (Si,O) layer, aluminium weight concentration is very low. The O/Si atomic ratio in the "electron dense" (Si,O) layer is around 3 so that this ratio is around 2 in the (Si,O,Al) layer. In these two layers, concentrations of Ca, P, K, Na and Mg are much lower than in the native bioactive glass and we do not represent them in the concentration profiles (figure 4).

*Bioactive glass periphery after 24 h of immersion in DMEM*

The inverted darkfield STEM image shows two parts of the bioactive glass particles after 24 h of immersion in DMEM (figure 5). We see again that the thin "electron dense" (Si,O) layer borders the glass particles all around their periphery; we can then ensure that the darkfield contrast corresponding to this thin layer is not due to a section artefact. Elemental weight concentration profiles of Si, O, Al, Ca and P reveal the presence of an another layer at the periphery of the bioactive glass (figure 6). We observe on top of the "electron dense" (Si,O) layer a (Ca,P) layer with a Ca/P atomic ratio around unity. After the formation of the (Ca,P) layer, the size of the "electron dense" (Si,O) layer decreases and this layer tends to disappear.

*Bioactive glass periphery after 96 h of immersion in DMEM*

Figure 7 presents an inverted darkfield STEM image of the bioactive glass particles after 96 h of immersion in DMEM. Elemental weight concentration profiles of Si, O, Al, Ca and P reveal the evolution of each zone from the center to the periphery of the bioactive glass (figure 8). The (Ca,P) layer grows in thickness according to time and the Ca/P atomic ratio increases (Ca/P ≈ 1.4). At this delay of exposure to the biological fluid, the "electron dense" (Si,O) layer disappeared. We just observe a (Si,O,Al) layer in which the aluminium weight concentration decreases as we approach its periphery. The O/Si atomic ratio is always around 2 in this layer.

*O/Si and Ca/P atomic ratios in each layer*

O/Si atomic ratios in the (Si,O,Al) layer and in the "electron dense" (Si,O) layer at each delay of exposure to DMEM are listed in table 1. The Ca/P ratio in the (Ca,P) layer is also listed in table 1. The O/Si ratio in the (Si,O,Al) layer is constant during exposure to DMEM and is slightly larger than 2. The O/Si atomic ratio in the "electron dense" (Si,O) layer is around 3 and then is significantly higher than in the (Si,O,Al) layer. It appears constant during the existence of this temporary layer. The Ca/P ratio in the (Ca,P) layer is close to unity at the beginning and increases according to time up to 1.4 at 96 h.

*Importance of the absorption correction process*

Inverted darkfield STEM images and EELS measurements of local relative specimen thickness show the heterogeneity of the mass thickness at the periphery of the bioactive glass. Thus we must take into account this heterogeneity in order to quantify EDXS spectra with accuracy. The error on the O/Si atomic ratio quantification could typically attain 40% if we neglected absorption corrections taking into account the determination of the local mass thickness.[9]

Discussion

In this work we analyse at the submicrometer scale the elemental weight concentrations of bioactive glass particles immersed into biological fluids by means of EDXS. These measurements are important for a better understanding of the different steps leading to the apatite layer formation at the bioactive glass periphery. This study requires some cautions, like the estimation of irradiation damages and X-rays absorption phenomenon. Absorption corrections are particularly important when light elements like oxygen are concerned.

As the glass matrix dissolves, various elements dispersed in the bioactive glass are free either to go into the solution or to combine with other elements in the bioactive glass that make up surface layers. The distribution of oxygen, silicon, aluminium, calcium and phosphorus differs between the glass particle center and the newly formed layers at the periphery.

Taking into account of previous knowledges [7, 19, 20] and of our present results, figure 9 synthetizes the different steps leading to the bioactivity mechanisms. It particularly points out the formation of these different layers and their evolution according to time of exposure to the biological solution. In the same time, we compare our steps with the five steps giving by Hench[7] : (i) rapid exchange of alkali elements with $H_3O^+$ from the surronding solution, (ii) breakage of Si-O-Si bonds and formation of Si-OH at the glass matrix solution interface, (iii) condensation and repolymerization of the $SiO_2$ surface layer, (iv) migration of $Ca^{2+}$ and $PO_4^{3-}$ groups through the $SiO_2$ layer and the formation of a $CaO-P_2O_5$ film on top of the $SiO_2$ surface layer, and (v) crystallization of the $CaO-P_2O_5$ film and formation of an apatite layer.

*Step 1 : Bioactive glass dissolution - leaching layer*

The dissolution of the bioactive glass results from breakdown of the silica network and the associated release of all elements within it (including silicon). Thus the weight concentrations of calcium, phosphorus, magnesium, sodium and potassium decrease strongly at the glass periphery.[10] Release of sodium and potassium ions into the solution is particularly strong and rapid which can be explained by the occupancy of the surface sites by either hydrogen ions or alkali ions. The release of aluminium is lower than others elements because it is firmly bond to silicon.[11] When the glass dissolves, most of $Al_2O_3$ is bonded to the unreleased part of silicon. Thus, the specific rate of aluminium released is smaller than the corresponding rate of silicon released because $Al_2O_3$ is a network former, not leached like silicon.[21] In the studied bioactive glass, addition of $Al_2O_3$ is used to control the solubility of the glass. The breakdown of the silica network and the associated release of all elements will move into the glass as the reaction proceeds (see figure 9). This first step agrees with the Hench'steps which corresponds to the exchange between $H_3O^+$ ions and alkali ions.

*Step 2 : formation of a (Si,O,Al) layer and an "electron dense" (Si,O) layer*

These reactions lead to the second step of dissolution which is the formation of a (Si,O,Al) layer at the periphery of the bioactive glass. We measure, in this work, an O/Si atomic ratio in this layer around 2 which can be interpreted by the condensation and the repolymerisation in a $SiO_2$ layer.[7] At the periphery of the (Si,O,Al) layer, we put in evidence the formation of a thin "electron dense" layer composed with silicon and oxygen. In this layer, we measure, in this work, an O/Si atomic ratio around 3 and the aluminium weight concentration tends to zero. Our measured O/Si atomic ratio can be interpreted by the formation of $Si(OH)_4$ groups at the periphery. The high release of alkali ions (step 1) permits the arrival of oxygen through an exchange with $H_3O^+$ ions from the solution. Water molecules in DMEM then or simultaneously react with the Si-O-Si bond to form additionnal Si-OH bonds.[22] This second step agrees with the Hench'steps which corresponds to the breakage of Si-O-Si bonds. We assist in the same time to the formation of theses $Si(OH)_4$ groups and to a repolymerization of theses groups under the $SiO_2$ form. Thus, a mixed $SiO_2$-$Si(OH)_4$ layer is built up which corresponds to the measured increase of the O/Si atomic ratio. This step corresponds to the thrid Hench'step (condensation and repolymerization of the $SiO_2$ surface layer) .

*Step 3 : formation of a (Ca,P) layer*

The (Si,O,Al) layer and the "electron dense" (Si,O) layer permit the diffusion of calcium and phosphorus up to the interface with the biological fluids which contains also calcium and phosphorus. The $Si(OH)_4$ groups can / could ???? permit a chemical bond with the calcium which might replace hydrogen in the hydroxyl bind with negatively charged oxygen species of the silica network. Then the calcium ion is with one free valence and can bind with an another oxygen in the silica network or preferentially with one phosphate group. A (Ca,P) layer is formed on top of the "electron dense" (Si,O) layer. In this (Ca,P) layer, the Ca/P atomic ratio is around unity.This first step agrees with the Hench'steps which corresponds to the migration of $Ca^{2+}$ and $PO_4^{3-}$ groups and the formation of a $CaO$-$P_2O_5$ film on top of the "electron dense" (Si,O) layer. The size of the "electron dense" (Si,O) layer decreases which traduces a diminution of $Si(OH)_4$ groups and this layer has generally disappeared before 96 h. This disappearance of this layer corresponds to the end of the third step given by Hench (repolymerization of the Si layer in $SiO_2$). This step finishs after the formation of the Ca-P layer. That can be confirmed by the results obtained at the delay of 24 h with the simultaneous presence of two layers : the "electron dense" (Si,O) layer and the (Ca,P) layer (figures 5 and 6).

*Step 4 : precipitation of the (Ca,P) layer*

The (Ca,P) layer grows in thickness according to time of exposure and its Ca/P atomic ratio increases. Afer 96 h, we measure a Ca/P atomic ratio around 1.4. Increasing delays will lead to a Ca/P atomic ratio reaching 1.7, demonstrating the precipitation of this layer into an apatite layer.[23, 24] This last step agrees with the last step given by Hench which corresponds to the crystallisation of the Ca-P layer into an apatite layer.

The short-term dissolution and physico-chemical reactions of the studied bioactive glass can be roughly described as two processes: a diffusion through two siliceous layers (namely (Si,O,Al) and "electron dense" (Si,O)); the occupancy of the surface sites by either hydrogen ions or alkali ions. As though these processes are clearly separable, they in fact occur simultaneously. Steps 1 and 2 occur very rapidly and simultaneously. In our case, the built-up of calcium phosphate occurs *in vitro* within a layer of silica gel like (a mixed $SiO_2$-$Si(OH)_4$). Apatite nucleation can be triggered with the presence of the silica gel. This can be attributed to the characteristics of the silica gel and to the presence of Si-OH bonds. At the periphery of the particles, silicon is probably arranged in the form of $Si(OH)_4$ groups, providing nucleation sites for the apatite formation.

In this bioactive glass, aluminium is used to reduce dissolution and its presence is very low in the silica gel. This element can not be accommodated in apatite precipitates. Apatite nucleation is retarded and even prevented completely, when aluminium concentration in the native bioactive glass reaches a high enough level. In our case, it appears that 2 % $Al_2O_3$ do not inhibit apatite formation at the bioactive glass particles periphery.

Different criteria can be used to predict the role of oxides added to the bioactive glass in the $SiO_2$-CaO-$P_2O_5$-$Na_2O$ system. The ionic field strength can be a good criterion to evaluate the role of MgO, $Al_2O_3$ and $K_2O$ in the studied bioactive glass. The ionic field strength ($z/r^2$, where z is the charge and r the ionic radius) is a measure of the electrostatic force that ion can exert upon neightboring oxygens. For field strength values starting from $z/r^2 \approx 5$ Å$^{-2}$ ions are expected to act as network formers[25, 26]. In our case, $Al^{3+}$ act as network forming ion and $Mg^{2+}$ could act as a network former but at a lower degree (Table 2). Other cations are network modifying ions.

The addition of a network former oxide like $Al_2O_3$ reduces the glass dissolution compare to other bioactive glasses[7]. This lower dissolution and the lower release in aluminum lead to the formation of the (Si,O,Al) layer. On the other hand, addition of network formers could significantly reduce the bioactivity of the material because the ability to form an apatite is reduced as the ionic field strength increases. This is not the case with 2% of $Al_2O_3$, in our study there is the formation of a (Ca,P) rich layer. Magnesium is not present in the (Si,O,Al) layer and do not play the role of a network former because its concentration in the glass (1% MgO) is to low and its field strength is under 5.

Conclusion

In the case of bioactive glasses that undergo dissolution, spatially resolved X-ray microanalysis is of great importance in evaluating short-term mechanisms of physico-chemical reactions at the interface between material and biological fluids. We use a dedicated quantification method to obtain local elemental weight concentrations which requires EELS measurements to correct sample X-rays absorption. This method permits to determine the weight concentration of light elements, like oxygen, and we were particularly interested by the measurements of the O/Si atomic ratios. The analysis of weight concentration variations at the bioactive glass particles / biological fluid gives information to a better understanding of the mechanisms of the apatite formation at the bioactive glass periphery.

In this paper, we present the different steps of the dissolution of bioactive glass particles in biological fluids and for the first time to our knowledge the presence at the periphery of a thin "electron dense" (Si,O) layer with an O/Si atomic ratio larger than in the others layers. This thin layer could then be characterised to the presence of $Si(OH)_4$ groups. The formation of the "electron dense" (Si,O) layer leads to the formation of a (Ca,P) layer on top due to the calcium ion ability to create a bond with the $Si(OH)_4$ group. The "electron dense" (Si,O) layer disappears as the (Ca,P) layer grows in thickness and finally precipitates in an apatite layer.

Captions

Figure 1 : Inverted darkfield STEM image of a native bioactive glass particle (R : resin, H : hole, BG : bioactive glass). The white line is the spectra acquisition line, where X corresponds to the position of the first recorded spectrum.

Figure 2 : Elemental weight concentration profiles across the periphery of native bioactive glass particles in the original state (see corresponding analysed area in figure 1).

Figure 3 : Inverted darkfield STEM image of a bioactive glass particle periphery at 12 h of exposure to biological solution (R : resin, a : (Si,O,Al) layer, b : "electron dense" (Si,O) layer). The white line is the spectra acquisition line, where X corresponds to the position of the first recorded spectrum.

Figure 4 : Si, O and Al weight concentration profiles across the periphery of bioactive glass particles at 12 h of exposure to biological solution (see corresponding analysed area in figure 3).

Figure 5 : Inverted darkfield STEM image of a bioactive glass particle periphery at 24 h of exposure to biological solution (R : resin, a : (Si,O,Al) layer, b : "electron dense" (Si,O) layer, c : (Ca,P) layer). The white line is the spectra acquisition line, where X corresponds to the position of the first recorded spectrum.

Figure 6 : Si, O, Al, Ca and P weight concentration profiles across the periphery of bioactive glass particles at 24 h of exposure to biological solution solution (see corresponding analysed area in figure 5).

Figure 7 : Inverted darkfield STEM image of a bioactive glass particle periphery at 96 h of exposure to biological solution (R : resin, a : (Si,O,Al) layer, b : (Ca,P) layer ). The white line is the spectra acquisition line, where X corresponds to the position of the first recorded spectrum.

Figure 8 : Si, O, Al, Ca and P weight concentration profiles across the periphery of bioactive glass particles at 96 h of exposure to biological solution (see corresponding analysed area in figure 7).

Figure 9 : Schematic view of the different steps leading to the bioactivity mechanisms of the bioactive glass particles during exposure to biological solution.

Table 1 : O/Si and Ca/P atomic ratios in (Si,O,Al), "electron dense"(Si,O) and (Ca,P) layers according to time of exposure in biological fluids.

Table 2 : Ionic field strengths ($z/r^2$) of the cations present in the bioactive glass[27].

Table 1

| Delay of exposure in DMEM | Layer | O/Si atomic ratios | Ca/P atomic ratios |
|---|---|---|---|
| After 12 h | (Si,O,Al) | $2.30 \pm 0.09$ | - |
| | "electron dense" (Si,O) | $2.90 \pm 0.07$ | - |
| After 24 h | (Si,O,Al) | $2.27 \pm 0.10$ | - |
| | "electron dense" (Si,O) | $2.96 \pm 0.13$ | - |
| | (Ca,P) | - | $1.03 \pm 0.15$ |
| After 96 h | (Si,O,Al) | $2.15 \pm 0.10$ | - |
| | (Ca,P) | - | $1.41 \pm 0.09$ |

Table 2

| | $Ca^{2+}$ | $Na^+$ | $Al^{3+}$ | $K^+$ | $Mg^{2+}$ |
|---|---|---|---|---|---|
| r (Å) Å | 1.12 | 1.02 | 0.54 | 1.38 | 0.72 |
| $z/r^2$ (Å$^{-2}$) | 1.59 | 0.96 | 10.29 | 0.52 | 3.85 |

Figure 1

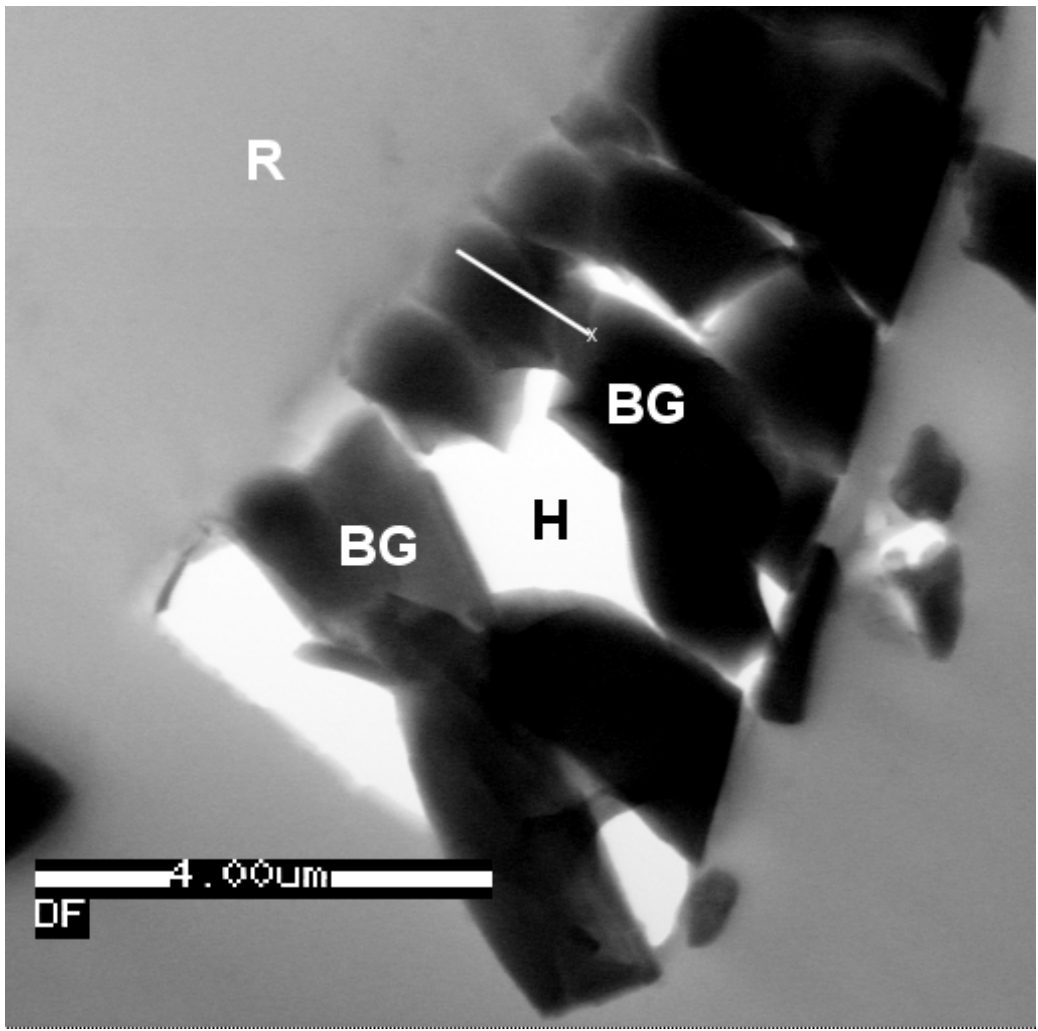

Figure 2

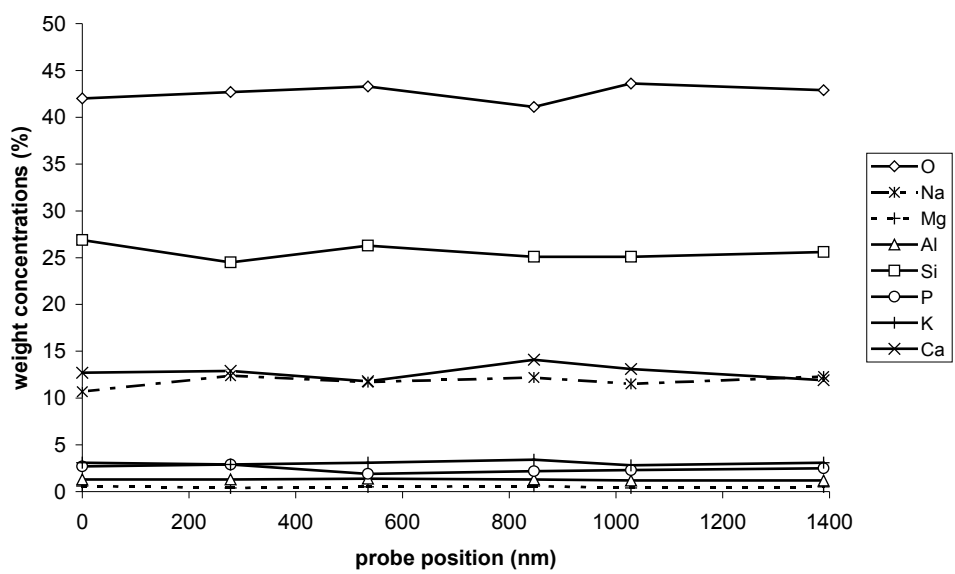

Figure 3

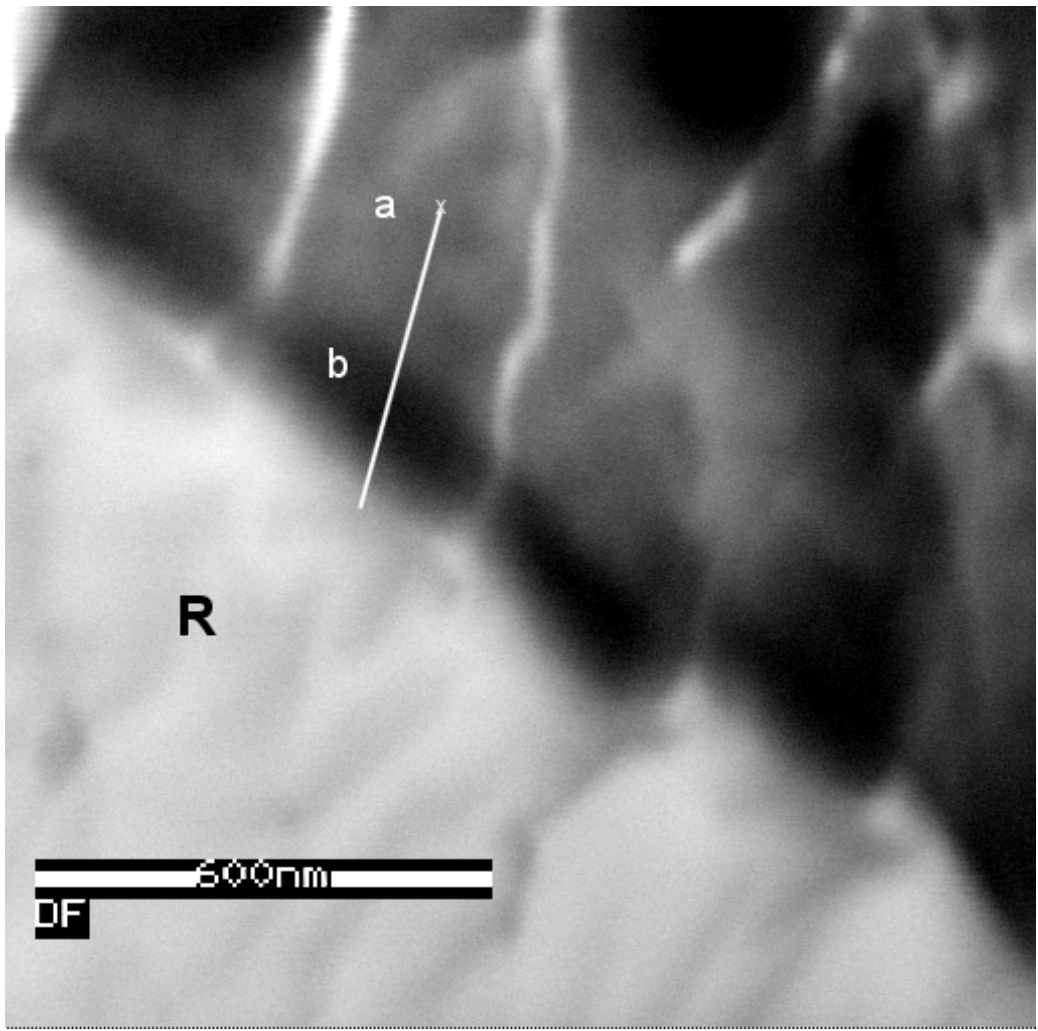

Figure 4

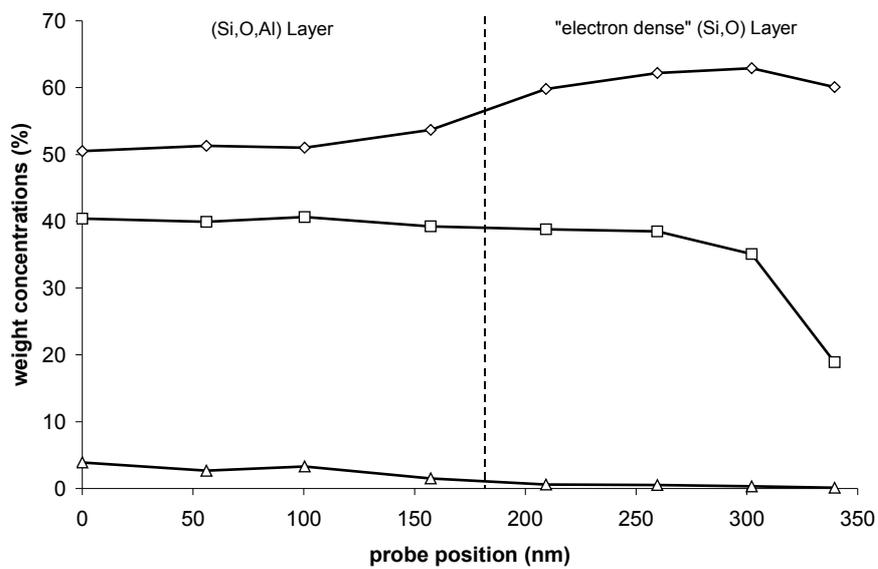

Figure 5

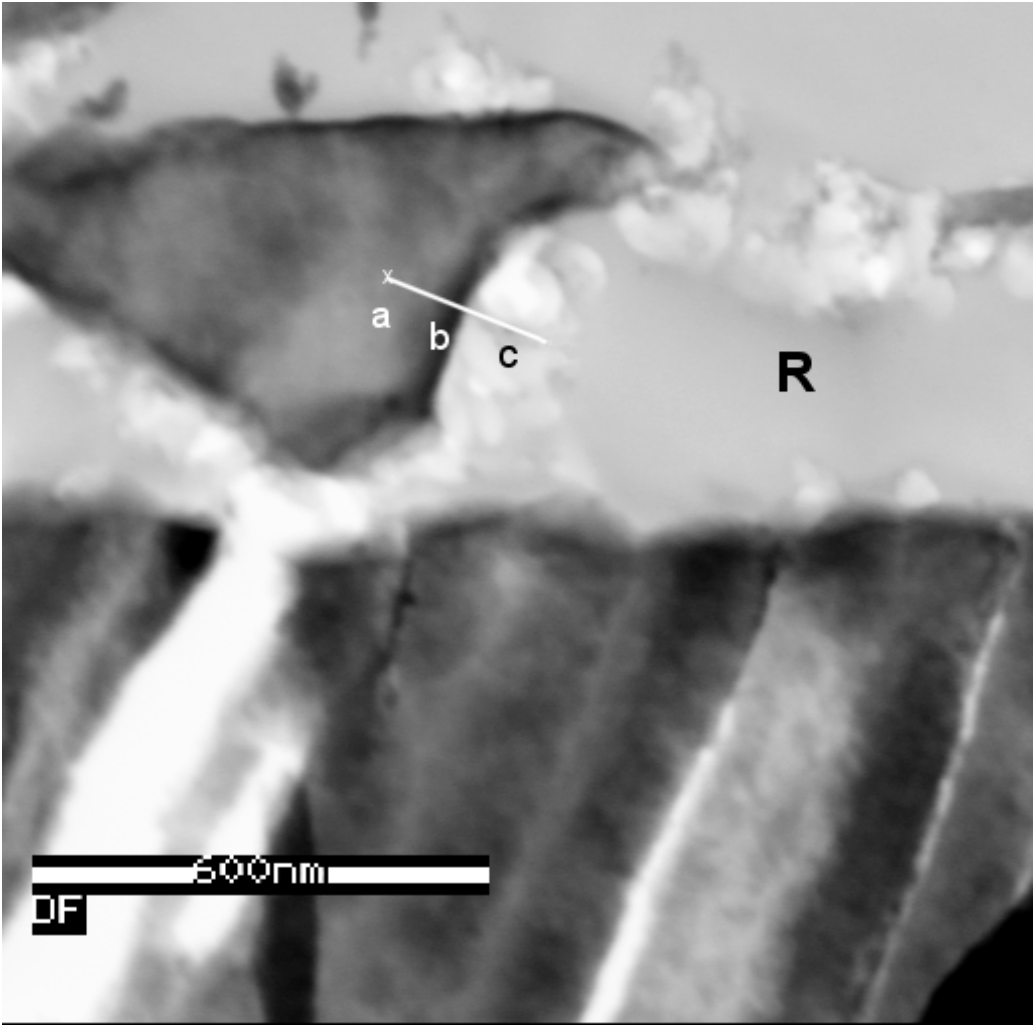

Figure 6

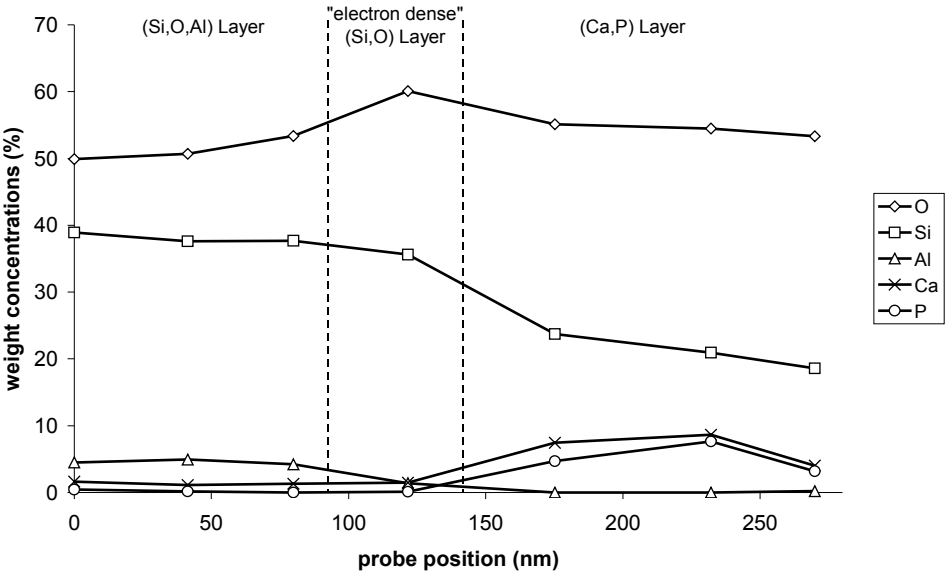

Figure 7

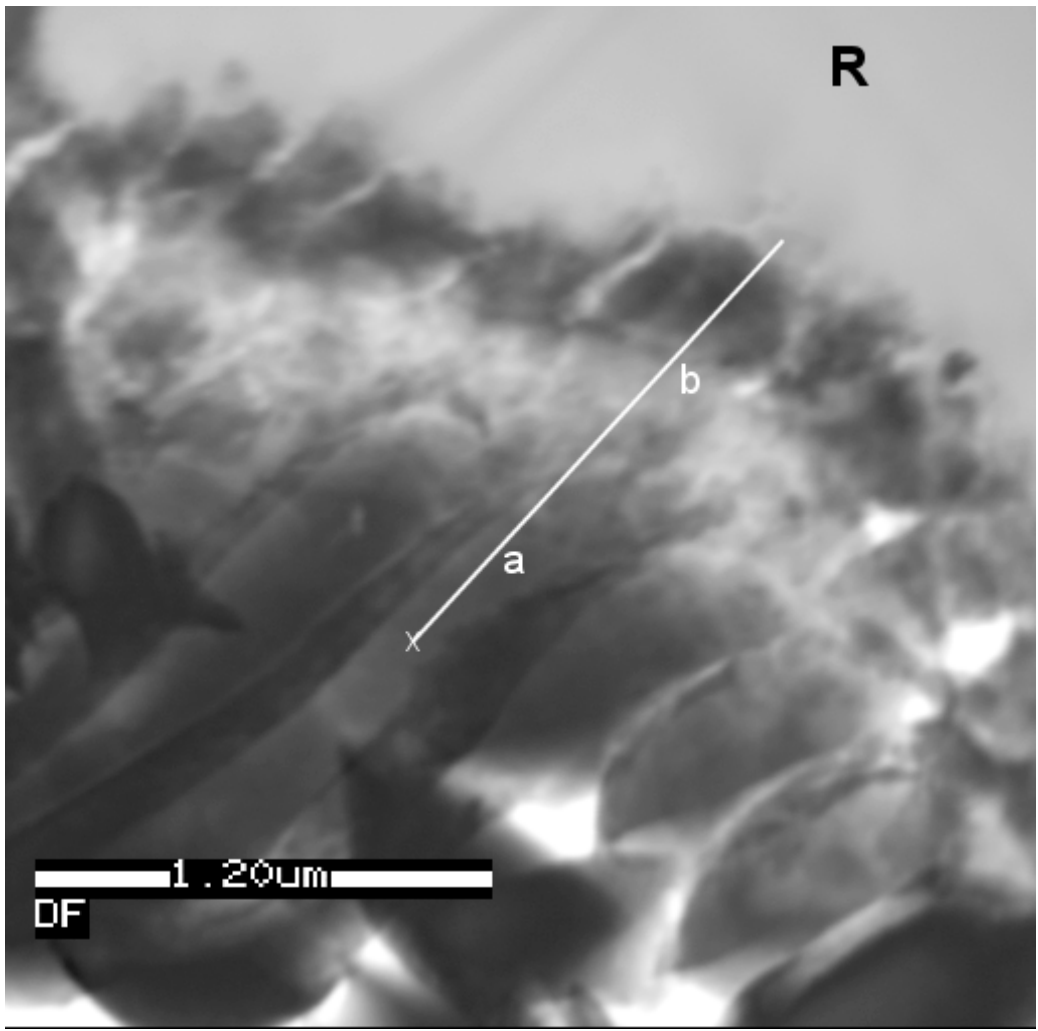

Figure 8

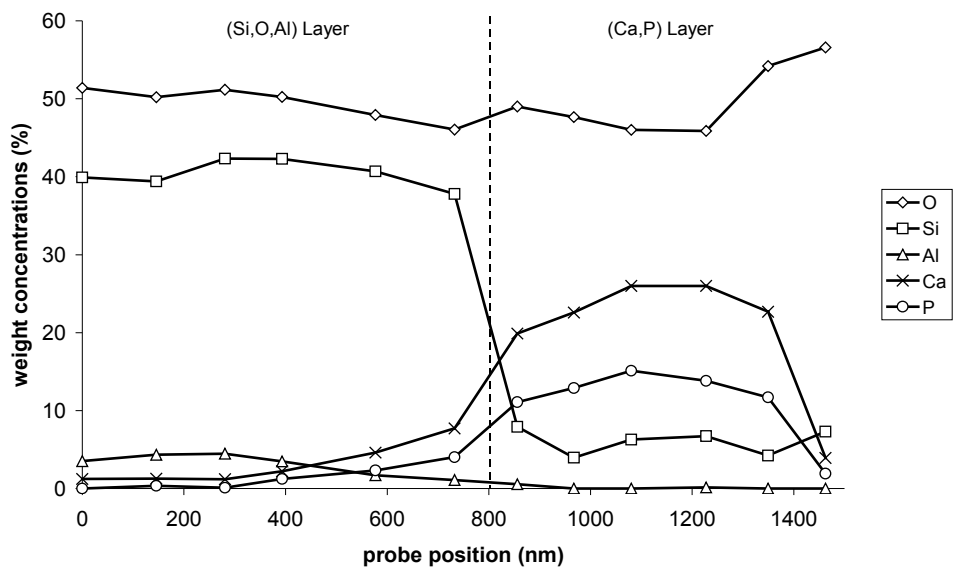

Figure 9

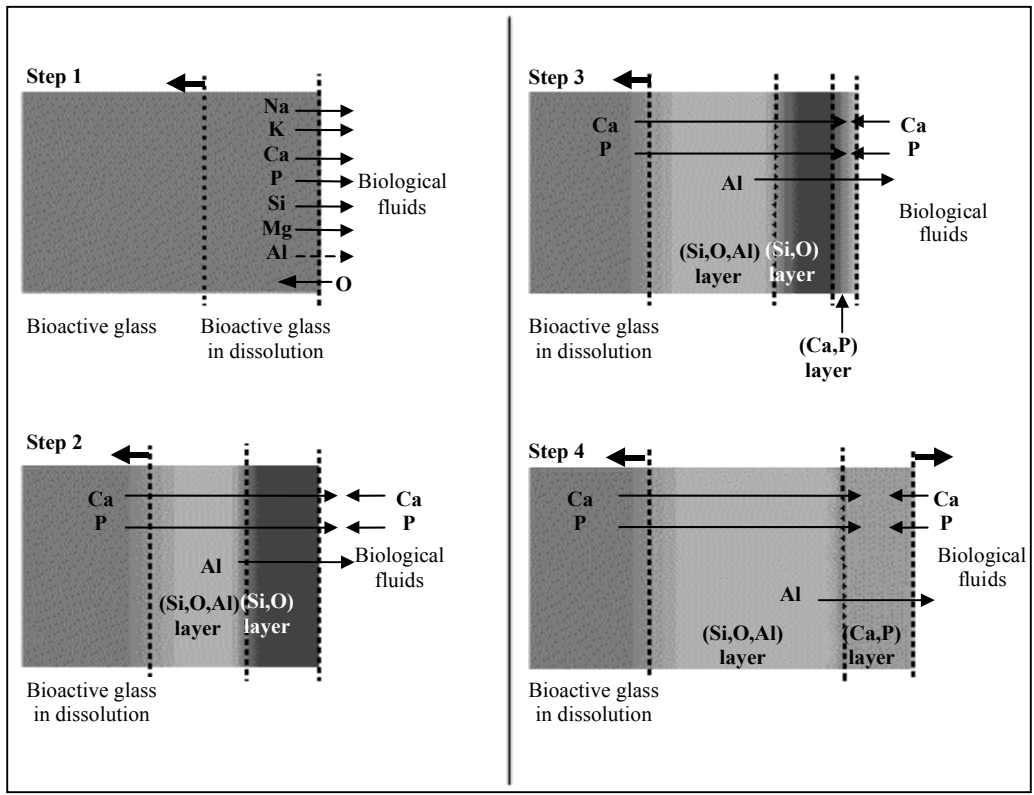